\documentstyle[epsf]{article} \catcode`\@=11
    \def\section{\@startsection{section}{1}{\z@}
    {-5.5ex plus -1ex minus -.5ex}{1.5ex plus.3ex}{\bf }}
    \def\subsection{\@startsection{subsection}{1}{\z@}
    {-3.5ex plus-1ex minus-.5ex}{1.5ex plus.3ex}{\sl }}
    
\begin{document}
     \vspace*{2.1cm}
{\Large\bf
Low temperature behavior of the thermopower \\[.2\baselineskip]
in disordered systems near the Anderson transition}
\vspace{.4cm}\newline{\bf   
C. Villagonzalo and R. A. R\"{o}mer
    }\vspace{.4cm}\newline\small 
Institut f\"{u}r Physik, Technische Universit\"{a}t, D-09107 Chemnitz, 
Germany 
    \vspace{.2cm}\newline 
Received 6 October 1998, revised 14 October 1998,
accepted 15 October 1998 by U. Eckern
    \vspace{.4cm}\newline\begin{minipage}[h]{\textwidth}\baselineskip=10pt
    {\bf  Abstract.}
We investigate the behavior of the thermoelectric power $S$ in disordered
systems close to the Anderson-type metal-insulator transition (MIT) at low
temperatures. In the literature, we find contradictory results for $S$. 
It is either argued to diverge or to remain a constant as the MIT
is approached. To resolve this dilemma, we
calculate the number density of electrons at the MIT in disordered systems
using an averaged density of states obtained by diagonalizing the
three-dimensional Anderson model of localization. From the number density
we obtain the temperature dependence of the chemical potential necessary
to solve for $S$. Without any additional approximation, we use the
Chester-Thellung-Kubo-Greenwood formulation and numerically obtain the
behavior of $S$ at low $T$ as the Anderson transition is approached from
the metallic side. We show that indeed $S$ does not diverge. 
    \end{minipage}\vspace{.4cm} \newline
    \newline\vspace{-.15cm} \normalsize

\section{Introduction}
In this paper, we study the low temperature behavior of the thermoelectric 
power $S$ in disordered systems 
near the Anderson-type metal-insulator transition (MIT). 
In the framework of linear response theory, $S$, commonly abbreviated  
as the {\em thermopower}, is the coefficient that 
relates the temperature gradient in an open circuit with the induced 
electric field.  In the metallic regime, the Sommerfeld theory states that 
$S$ is  directly proportional to the negative temperature $-T$ \cite{ashcroft}. 
But at a disordered-induced MIT, such as the Anderson transition in three 
dimensions (3D) \cite{anderson}, it is still not a settled issue how $S$ 
behaves at low $T$. Theoretical  studies have either claimed that it diverges 
\cite{imry}, or  that it remains a constant \cite{enderby} as 
the MIT is approached  from the metallic side at low $T$. 
Moreover, comparing the results of the latter theory with that of 
experiments conducted on doped semiconductors \cite{lakner} 
and on amorphous alloys \cite{lauinger} shows that measurements of $S$  
are two orders of magnitude  higher than those predicted in theory. 
Thus, it is of great interest to investigate the behavior of $S$ at low $T$ 
near the Anderson-type MIT. Here, for simplicity, we consider only the 
diffusion part of $S$, that is, we consider only the electronic 
contribution and neglect any possible
electron-phonon interactions. In addition to $S$, we shall also compute 
thermal transport properties such as the thermal conductivity $K$ and 
the Lorenz number $L_{0}$.

\section{Theoretical background}
The derivation of the thermopower is based on the kinetic coefficients 
of the Chester-Thellung-Kubo-Greenwood formulation of linear response 
\cite{ctkg},
\begin{equation}
L_{ij}=  (-1)^{i+j}\int_{-\infty}^{\infty} A(E) [E-\mu(T)]^{i+j-2} 
\left(- \frac{\partial f(E,\mu,T)}{\partial E} \right) dE \;\;\;\;\;\;\;
i, j = 1, 2 \label{coef}
\end{equation}
where $E$ is the energy, $A(E)$ contains all the system-dependent features,
$\mu(T)$ is the chemical potential, $f(E,\mu,T)=1/[1+\exp([E-\mu(T)]/k_{B}T)]$ is 
the Fermi function, and  $k_{B}$ is the Boltzmann constant.
These coefficients relate the electric field $\varepsilon$, concentration 
gradient $\nabla\mu$ and temperature gradient $\nabla{T}$
to the expectation values of the 
induced electric $\langle{\bf j}_{1}\rangle$  and heat
$\langle{\bf j}_{2}\rangle$ current densities 
\begin{equation}
\langle{\bf j}_{i}\rangle = |e|^{-i}\left[-L_{i1}({\bf \nabla}\mu-|e|
\varepsilon)-L_{i2}T^{-1}{\bf \nabla}T\right]\;. 
\end{equation}
where $e$ is the electron charge.
Measured under the assumption  that there is no electric current and 
concentration gradient, the thermopower is thus given as
\begin{equation}
S = \frac{L_{12}}{|e|TL_{11}}\;. \label{tp}
\end{equation}
The Anderson transition is then incorporated into the measurement 
of $S$ by setting the function $A(E)$ in the coefficient $L_{ij}$ as 
proportional to the critical behavior of the d.c. conductivity $\sigma$ 
at the MIT, that is,
\begin{equation}
A(E)=\left\{ \begin{array}{cl}
\alpha|E-E_{C}|^{\nu} & E \geq E_{C} \\
0 & E<E_{C}\end{array}\right. \label{dc_cond}
\end{equation}
where $\alpha$ is a constant, $\nu$ is the conductivity index and 
$E_{C}$ is the mobility edge. With this assignment the coefficient 
$L_{11}$ is simply $\sigma$. Furthermore, since $K$ is the 
coefficient that relates the temperature gradient to the induced 
heat current, it's  low temperature behavior at the MIT
can be determined in a similar manner from $\langle{\bf j}_{2}\rangle$ 
with the assumption that there are no particle currents, and using the 
Anderson transition form of $A(E)$ as given above. 
Then the Lorenz number $L_{0}=(e/k_{B})^{2} \sigma/KT$
quickly follows. Thus, 
the low $T$ behavior of $S$, $K$ and $L_{0}$ at the Anderson 
transition  follows easily after obtaining
the kinetic coefficients, Eq.\,(\ref{coef}). 

\subsection{Divergent thermopower}
A divergent $S$ at the Anderson transition $E=E_{C}$ is obtained 
if one uses the Sommerfeld expansion to get the low-$T$ leading 
contribution to $L_{ij}$ \cite{imry}. This method assumes that the 
chemical potential $\mu$ is equal to the Fermi energy $E_{F}$ even for finite $T$.
However, $\mu=E_{F}$  only at $T=0$ \cite{ashcroft}. 
A more serious approximation  of the Sommerfeld expansion 
is the assumption that  $A(E)$ is 
a smoothly varying function at $E=E_{C}$. 
This is not the case at the Anderson transition, as can be 
readily seen in Eq.\,(\ref{dc_cond}).

\subsection{Fixed-point thermopower}
The approach proposed by Enderby and Barnes \cite{enderby}
evaluates the kinetic coefficients at $\mu=E_{C}$ for finite $T$, and afterwards
the limit  $T \rightarrow 0$ is taken. They find that 
the thermopower is a constant at the mobility edge for $T \rightarrow 0$, 
and is given by
\begin{equation}
S = - \frac{k_{B}}{|e|} \frac{(\nu+1)}{\nu} \frac{I_{\nu+1}}{I_{\nu}}
\label{tpEB}
\end{equation}
where $I_{1}=\ln2$,  $I_{\nu}= (1-2^{(1-\nu)})\Gamma(\nu)\zeta(\nu)$ for
${\mathrm Re}[\nu]>0,\;\nu\neq1$, with $\Gamma(\nu)$ and $\zeta(\nu)$  the 
usual gamma and Riemman zeta functions. Hence, $S$ solely depends 
on $\nu$.

\section{Calculation of the temperature dependent thermopower}
One can determine the temperature dependence of the 
thermopower if one knows how $\mu$ varies with $T$. 
This information can be obtained from the number
density $n$ of electrons at the MIT. In general, for any set of noninteracting 
electrons, the number density is defined as
\begin{equation}
n(\mu,T) = \int_{-\infty}^{\infty}dE g(E) f(E,\mu,T)\;, \label{numden}
\end{equation}
where $g(E)$ is the density of energy levels per unit volume.
Using the above equation, we numerically calculate $n$
using an averaged density of states $g(E)$ obtained by diagonalizing the 
Anderson model of localization. 
Earlier, we determined the averaged density of states
for a 3D isotropic Anderson model with disorder $W=12$ \cite{milde}. 
Note that since our objective is to  compare our theoretical results 
for $S$ with experimental measurements, such as those from amorphous alloys, 
the hopping parameter $t$ is  of the order 
of 1 eV. Hence, we have expressed all energy units in terms of $t$ unless 
otherwise specified.  We have selected the value of $W$ to be strong 
enough, such that we do not have 
singularities in the density of states. Yet, it should not be too strong, 
i.e. too close  to the critical disorder. 
For this particular value of $W$, the value of $E_{C}$ is
approximately  $-7.5$, according to the mobility edge trajectory $E_{C}(W)$
calculated in Ref.\,\cite{schreiber}. 
The conductivity index $\nu$ is $\approx1.3$, 
according to a current numerical estimate \cite{kramer}.
Then we integrate the density of states for $E\leq E_{F}$ to obtain
the corresponding value of $n$ for a given value of $E_{F}$ at $T=0$. 
Keeping $n$ fixed at this value, we  vary $T$ in Eq.\,(\ref{numden}) 
and  numerically  determine the variation of $\mu$. 
Using this information in Eq.\,(\ref{coef}), we
solve for $L_{ij}$. It is then straightforward to determine
$S$ for a particular value of $E_{C}$ from Eq.\,(\ref{tp}). 
 
\section{Results and discussion}

\begin{figure}[th]
  \epsfysize=6cm 
  \centerline{\epsfbox{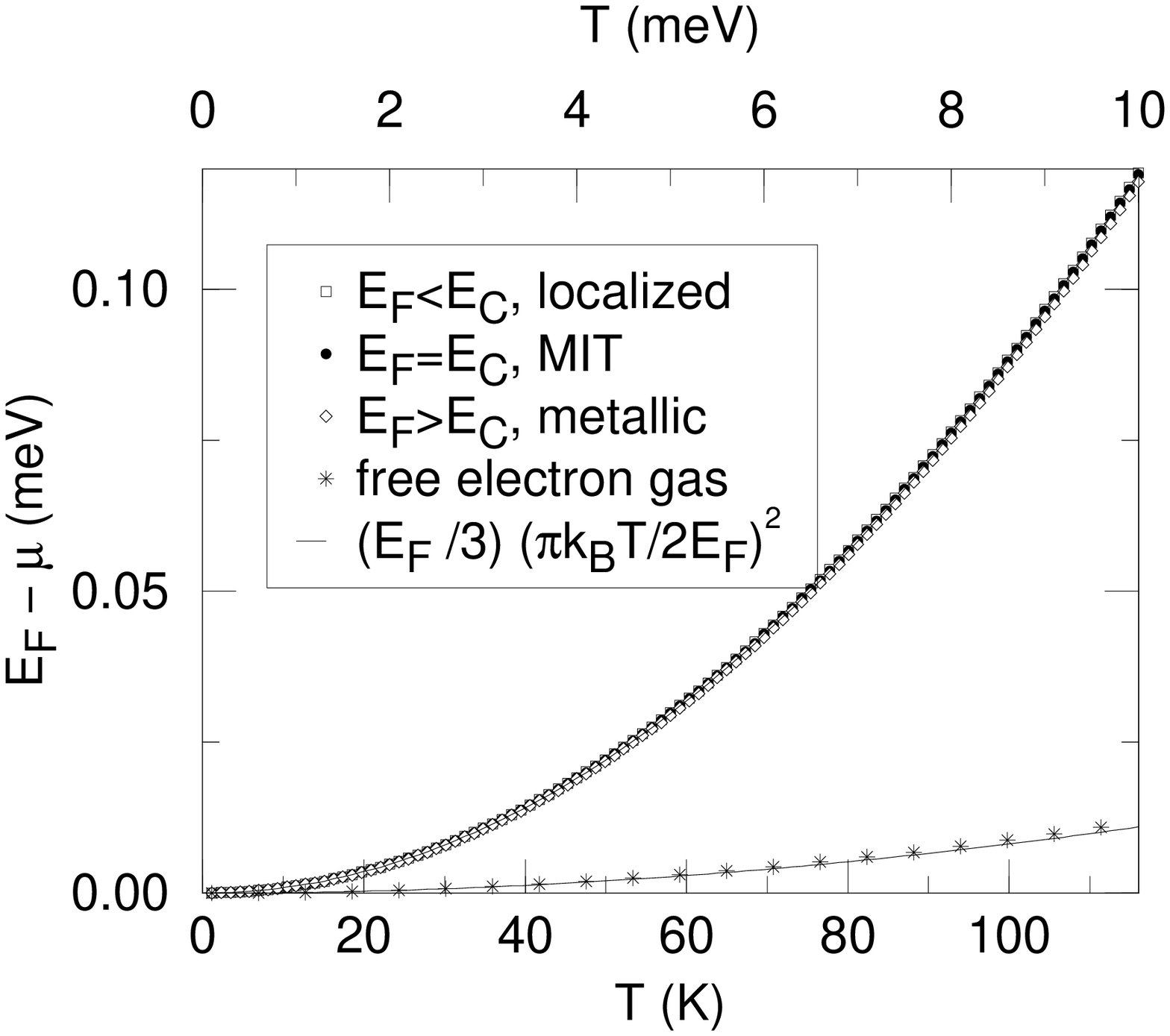}} 
  \vskip 0.25cm 
  \epsfysize=4.2cm 
  \centerline{\epsfbox{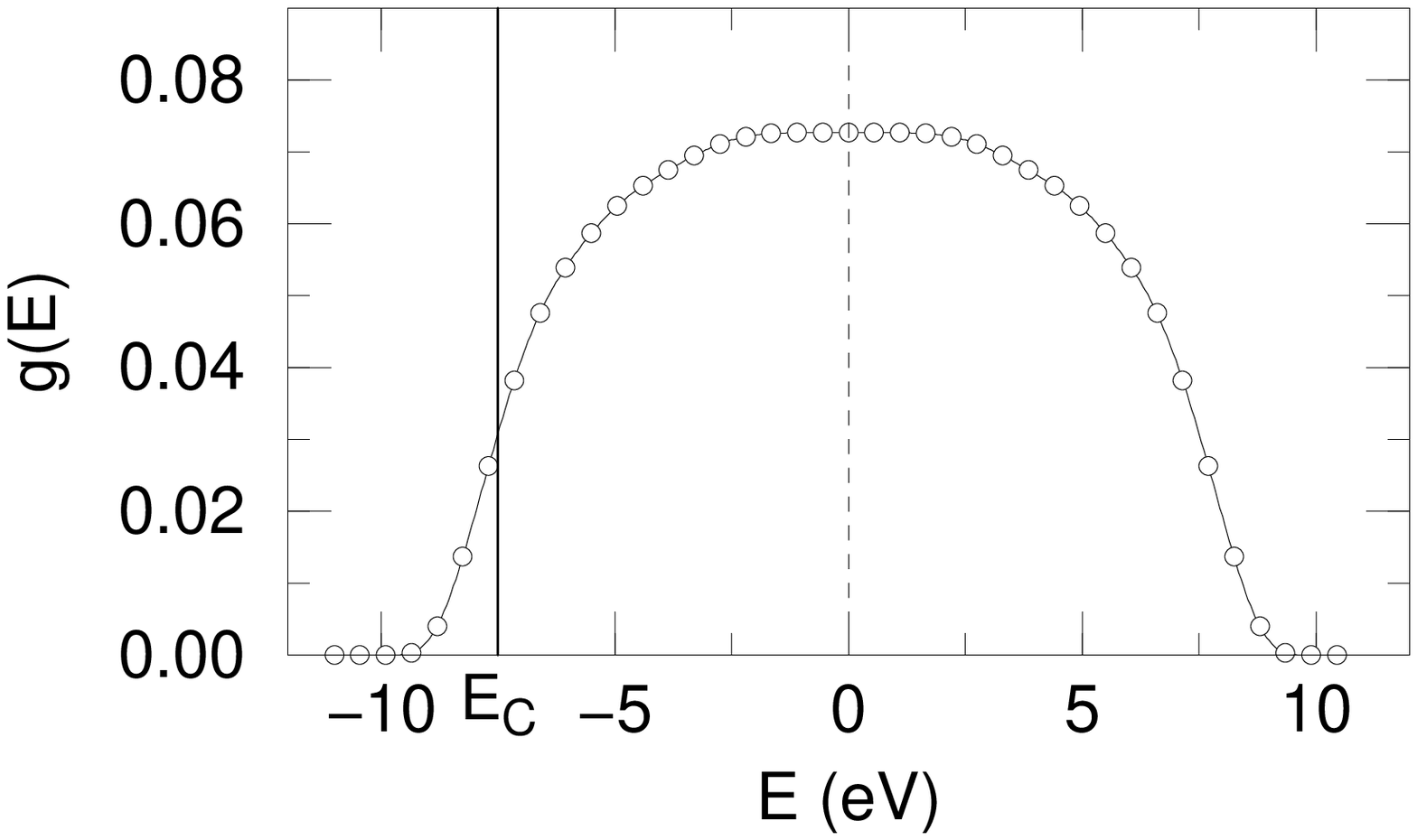}}
 \caption{\protect\small 
 Top: The low $T$ behavior of $\mu$.
 Near the MIT, $\mu(T)$ is similar in both the localized and the 
 metallic regions.
 Bottom: The averaged density of states of a 3D isotropic 
 Anderson model with $W=12$. For clarity only every 10th 
 data point is marked by a symbol ($\circ$).}
 \label{mu}
\end{figure}

In Fig.\,\ref{mu}, the temperature dependendence of the chemical potential is 
shown together with the averaged density of states from which it was measured.
Note that from this smooth density of states, we obtain a
$T$ dependence of $\mu$ which barely changes when one selects $E_{F}$
in the metallic or the localized region. 
However, its slope changes much faster as compared to 
the chemical potential from a free electron gas as shown in Fig.\,1.
Note that this free electron 
result was also similarly obtained from the same expression for $n$ given in 
Eq.\,(\ref{numden}), but using the Sommerfeld expansion in order to obtain $\mu$. 

\begin{figure}[th]
  \epsfysize=6cm \centerline{\epsfbox{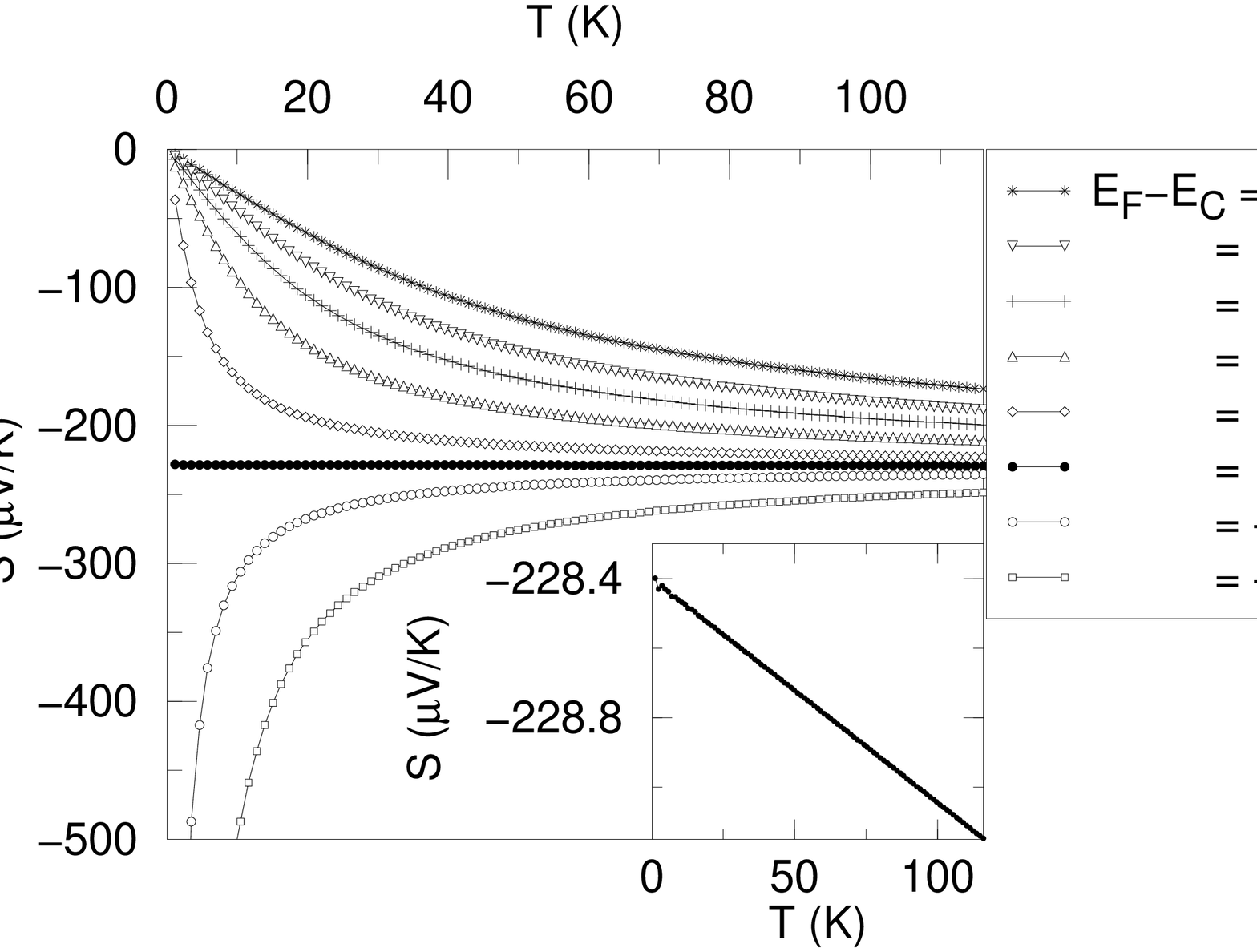}}
  \vskip 0.25cm 
  \epsfysize=6cm  \centerline{\epsfbox{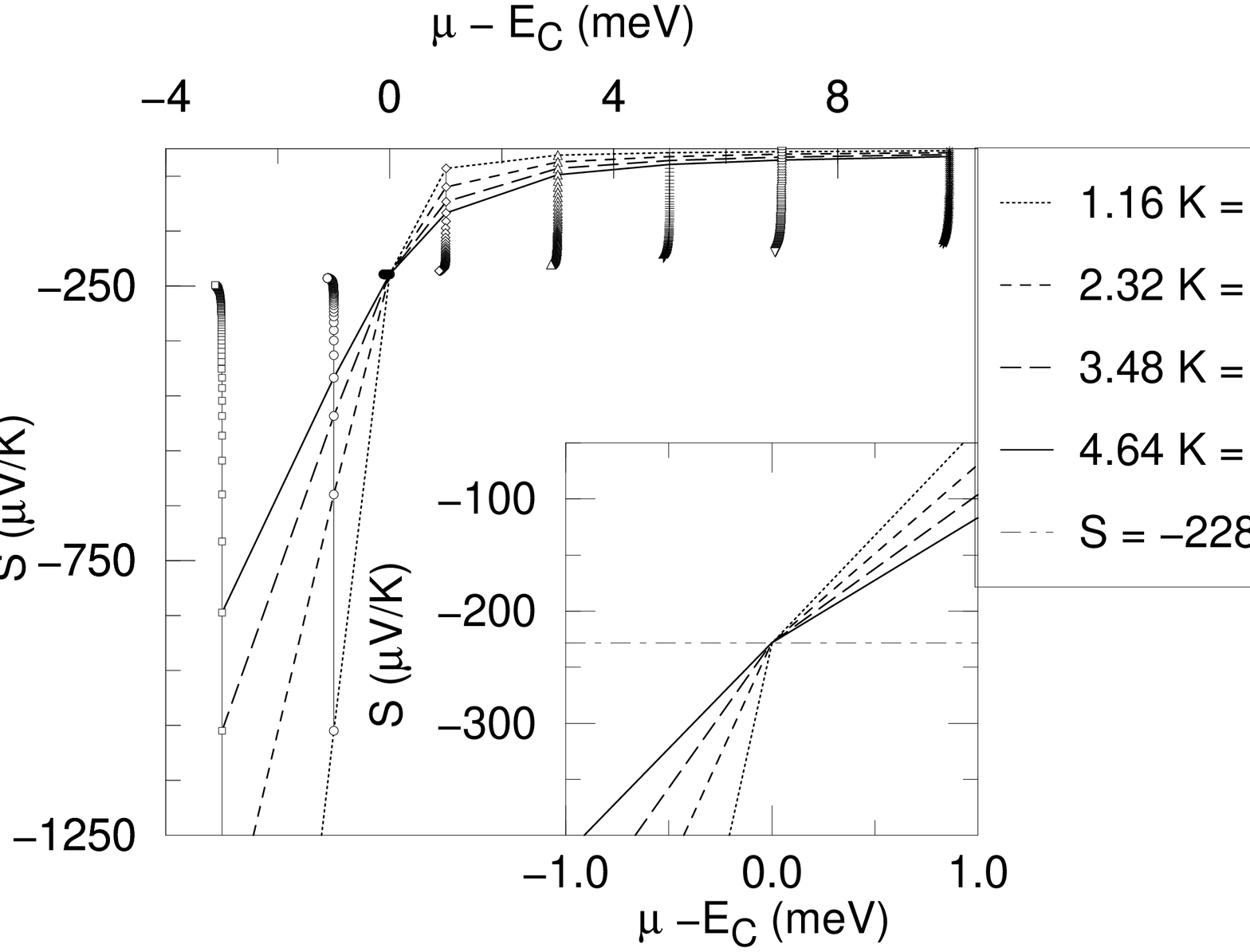}}
\caption{\protect \small
  Top: The low $T$ behavior of $S$.
  Note that $S$ does not diverge at the MIT as $T\rightarrow0$. Bottom: 
  Same data as in the top panel, plotted
  with respect to $\mu$ for different Fermi energies.
  The lines connect isotherms of $S$. 
  As shown in the inset, $S$ is a fixed point at the MIT.}
\label{SvsTmu}
\end{figure}

Next, Fig.\,\ref{SvsTmu} shows our thermopower measurements. 
The curves at the top of  Fig.\,\ref{SvsTmu} clearly show the MIT, 
the dividing line between the metallic ($E_{F}>E_{C}$) 
and localized ($E_{F}<E_{C}$) regions. 
As $T\rightarrow0$, $S$ gets more negative in the localized region, while 
$S\rightarrow0$ in the metallic region. As we move 
further away from the MIT towards the metallic region  at low $T$, $S$ behaves 
as expected from the Sommerfeld theory, that is, linearly proportional to $-T$. 
This indicate nonzero values of $\sigma(E_{F})$ confirming the metallic nature 
in this energy region. 
More importantly, we see that $S$ is a constant at the MIT, $E_{F}=E_{C}$.  
As $T\rightarrow0$, it approaches the value -228.4 ${\mu}V/K$. 
This value agrees with the $T$-independent value for  $\nu=1.3$ as predicted
by Eq.\,(\ref{tpEB}). 
At the MIT, a negative $S$ value of the order of 
hundreds of ${\mu}V/K$ has never been experimentally observed
to the best of our knowledge. 
To see the $T$-independence of $S$ at the MIT, we refer 
to the bottom of Fig.\,\ref{SvsTmu}. 
Here we show the behavior of $S$ at different Fermi energies
for different temperatures. It is clearly demonstrated in the inset that 
for different values of $T$, $S$ is a fixed point at the MIT ($\mu=E_{C}$) 
verifying what Enderby and Barnes had previously concluded \cite{enderby}. 

Similarly, we have studied the other thermal transport properties $K$ 
and $\sigma$. Our preliminary investigation shows that $K\rightarrow0$ as
$T\rightarrow0$ at any energy region. Furthermore, in the metallic phase, 
$L_{0}$ approaches the  value $\pi^{2}/3$ which according to the law of Wiedemann 
and Franz is a universal value for all metals (see for example 
Refs.\,\cite{ashcroft,ctkg}).  At the MIT, however, $L_{0}$ has a value dependent 
only on the conductivity index $\nu$.
Detailed results of these transport properties will be discussed elsewhere. 

\section{Conclusions}
In this work we have studied the low temperature behavior of the thermoelectric
power for the 3D isotropic Anderson model close to the MIT. We have numerically 
obtained the temperature dependence of the chemical potential necessary to solve 
for $S$ from the general expression of the number density for any set of 
noninteracting  electrons. We have shown that $\mu(T)$ is very similar 
regardless which energy region close to the MIT one considers. Using this result 
and the Chester-Thellung-Kubo-Greenwood formulation, our calculations
yield a sharp contrast of the $S$ behavior between 
metallic and localized regions clearly outlining the MIT. 
Finally, as the MIT is approached from the metallic side $S$ is a fixed point.
As $T\rightarrow0$ at the MIT, $S$ approaches the fixed-point value predicted by 
Enderby and Barnes which for $\nu=1.3$ is $S = -228.4\;{\mu}V/K$.
Therefore, we have established  that as the MIT is approached at low $T$ 
the thermopower does not diverge but remains a constant. Its fixed-point 
value depends only on the critical behavior of $\sigma$.
How $S$ behaves for varying degrees of disorder is a subject of further 
investigation. 
\\ 

\baselineskip=10pt
{\small
We thank T. Vojta for helpful discussions. 
We also gratefully acknowledge financial support by the DFG through
Sonderforschungsbereich 393.}


\begin{thebibliography}{99} 

\itemsep-3.5pt 
\small\frenchspacing
\baselineskip=10pt

\bibitem{ashcroft} N. W. Ashcroft, N. D. Mermin, Solid State Physics,
Saunders College, New York, 1976 
\bibitem{anderson} P. W. Anderson, Phys. Rev. {\bf 109} (1958) 1492
\bibitem{imry} U. Sivan, Y. Imry, Phys. Rev. B {\bf 33} (1986) 551;
C. Castellani, C. Di Castro, M. Grilli, 
G. Strinati, Phys. Rev. B {\bf 37} (1988) 6663 
\bibitem{enderby}J. E. Enderby, A.C. Barnes, Phys. Rev. B {\bf 49} 
(1994) 5062 
\bibitem{lakner} M. Lakner, H. v. L\"{o}hneysen, 
Phys. Rev. Letters {\bf 70} (1993) 3475
\bibitem{lauinger} C. Lauinger, F. Baumann, J. Phys.: Condens. Matter 
{\bf 7} (1995) 1305 
\bibitem{ctkg} G. V. Chester, A. Thellung, Proc. Phys. Soc. {\bf 77}
(1961) 1005;  R. Kubo, J. Phys. Soc. Japan {\bf 12} (1957) 570; 
D. A. Greenwood, Proc. Phys. Soc. {\bf 71} (1958) 585 
\bibitem{milde} F. Milde, R. A. R\"{o}mer, M. Schreiber, 
Phys. Rev. B {\bf 55} (1997) 9463
\bibitem{schreiber} H. Grussbach, M. Schreiber, Phys. Rev. B {\bf 51}
(1995) 663
\bibitem{kramer} B. Kramer, A. MacKinnon, Rep. Prog. Phys. {\bf 56} 
(1993) 1469 
    \end{thebibliography}
    \end{document}